\documentclass[sigconf, 9pt]{acmart}

\usepackage[para,online,flushleft]{threeparttable}
\usepackage{paralist}
\usepackage{mdwlist}
\usepackage{tikz}
\usepackage{amsmath}
\usepackage{xspace}
\usepackage{graphicx}
\usepackage[htt]{hyphenat}
\usepackage{color}
\usepackage{svg}
\usepackage{listings}
\usepackage{flushend}
\usepackage{colortbl}
\usepackage{todonotes}
\usepackage{pifont}
\usepackage[most]{tcolorbox} 
\usepackage{booktabs}
\usepackage{multirow}
\usepackage{enumitem}
\usepackage{pifont}
\usepackage{caption}
\usepackage{subcaption}
\usepackage{soul}
\usepackage{url}
\usepackage{breakurl}
\usepackage{comment}
\usepackage{stackengine}
\usepackage{mathtools}
\usepackage{adjustbox}
\usepackage[nodisplayskipstretch]{setspace}
\usepackage[linesnumbered,ruled,vlined]{algorithm2e}
\usepackage{algpseudocode}
\usepackage{hyperref}

\usepackage{xcolor}

\definecolor{lightgray}{gray}{0.9}
\definecolor{tablegreen}{rgb}{0.65,0.95,0.82}
\definecolor{codegreen}{rgb}{0,0.6,0}
\definecolor{codegray}{rgb}{0.5,0.5,0.5}
\definecolor{codepurple}{rgb}{0.58,0,0.82}
\definecolor{backcolour}{rgb}{0.95,0.95,0.92}

\usepackage{filecontents}

\definecolor{main}{HTML}{5989cf}
\definecolor{sub}{HTML}{cde4ff}

\newcommand{\squishlist} 
{
    \begin{list}{$\bullet$}
    {
        \setlength{\itemsep}{0pt}      \setlength{\parsep}{3pt}
        \setlength{\topsep}{3pt}       \setlength{\partopsep}{0pt}
        \setlength{\leftmargin}{1.5em} \setlength{\labelwidth}{1em}
        \setlength{\labelsep}{0.5em}
    }
}

\newcommand{\squishend}
{
    \end{list}
}

\newcommand{\cmark}{\ding{51}}%
\newcommand{\xmark}{\ding{55}}%

\lstdefinestyle{mystyle}{
    backgroundcolor=\color{backcolour},   
    commentstyle=\color{codegreen},
    keywordstyle=\color{magenta},
    numberstyle=\tiny\color{codegray},
    stringstyle=\color{codepurple},
    basicstyle=\ttfamily\footnotesize,
    breakatwhitespace=false,         
    breaklines=true,                 
    captionpos=b,                    
    keepspaces=true,                 
    numbers=left,                    
    numbersep=5pt,                  
    showspaces=false,                
    showstringspaces=false,
    showtabs=false,                  
    tabsize=1,
    frame=tb, 
    frameround=tttt,
    linewidth=1\linewidth, 
    columns=fullflexible, 
}

\newbool{IsPrintComment}
\boolfalse{IsPrintComment}

\newbool{IsPrintContent}
\boolfalse{IsPrintContent}

\newcommand{\chirag}[1]
{
    \ifbool{IsPrintComment}
    {
        {\color{red} TODO: #1}
    }{}
}

\newcommand{\igc}[1]{
\ifbool{IsPrintComment}{%
    {\color{blue} IG: #1}
}{}}

\newcommand{\igt}[1]{
\ifbool{IsPrintContent}{%
    {\color{blue} #1}
}{}}

\iffalse
\newcommand{\sj}[1]{\todo[inline,color=brown!40]{Saurabh: #1}}
\newcommand{\ig}[1]{\todo[inline,color=green!40]{Indy: #1}}
\newcommand{\sarthak}[1]{\todo[inline,color=blue!40]{Sarthak: #1}}
\newcommand{\cs}[1]{\todo[inline,color=cyan!40]{Chirag: #1}}
\newcommand{\cn}[1]{\todo[inline,color=orange!40]{Chandra: #1}}
\newcommand{\ls}[1]{\todo[inline,color=violet!40]{Laura: #1}}
\newcommand{\tv}[1]{\todo[inline,color=yellow!40]{Thrivikram: #1}}
\newcommand{\hf}[1]{\todo[inline,color=purple!40]{Hubertus: #1}}

\else
\newcommand{\sj}[1]{}
\newcommand{\ig}[1]{}
\newcommand{\sarthak}[1]{}
\newcommand{\cs}[1]{}
\newcommand{\cn}[1]{}
\newcommand{\ls}[1]{}
\newcommand{\tv}[1]{}
\newcommand{\hf}[1]{}

\fi

\usepackage{kantlipsum,setspace}

\usepackage{cleveref}
\usepackage[font={it,small},labelsep=colon, belowskip=-1.0pt, aboveskip=0.3pt]{caption}

\usepackage{xpatch}
\newcommand{\cpulimits}[0]{\textsc{CPU-Limits}\xspace}
\newcommand{\climits}[0]{\textsc{c.limits}\xspace}
\newcommand{\climit}[0]{\textsc{c.limit}\xspace}
\newcommand{\clim}[0]{\textsc{c.lim}\xspace}

\newcommand{\cpurequests}[0]{\textsc{CPU-Requests}\xspace}
\newcommand{\crequest}[0]{\textsc{c.request}\xspace}
\newcommand{\crequests}[0]{\textsc{c.requests}\xspace}
\newcommand{\creq}[0]{\textsc{c.req}\xspace}

\newcommand{\cpuutil}[0]{\textsc{CPU.Util}\xspace}

\newcommand{\kubernetes}[0]{Kubernetes\xspace}
\newcommand{\ks}[0]{K8s\xspace}

\newcommand{\hpa}[0]{\texttt{HPA}\xspace}
\newcommand{\cputhresh}[0]{\textsc{CPU.Thres}\xspace}

\AtBeginDocument{%
  }

\setcopyright{acmlicensed}
\copyrightyear{2025}
\acmYear{2025}
\acmDOI{XXXXXXX.XXXXXXX}
\acmConference[Arxiv]{Make sure to enter the correct
  conference title from your rights confirmation email}{Oct 2025}{arxiv.org}
\acmISBN{978-1-4503-XXXX-X/2025/06}

\begin{document}


\date{}




\title{CPU-Limits kill Performance: Time to rethink Resource Control
}








\newcommand{\uiuc}{UIUC, Illinois, USA}
\newcommand{\ibm}{IBM Research, New York, USA}

\author{Chirag Shetty}
\authornote{Corresponding author. Email: cshetty2@illinois.edu\\
UIUC: University of Illinois Urbana-Champaign, Illinois, USA}
\affiliation{%
  \institution{\uiuc}
  \country{}
}

\author{Sarthak Chakraborty}
\affiliation{%
  \institution{\uiuc}
  \country{}
}

\author{Hubertus Franke}
\affiliation{%
  \institution{\ibm}
  \country{}
}

\author{Larisa Shwartz}
\affiliation{%
  \institution{\ibm}
  \country{}
}

\author{Chandra Narayanaswami}
\affiliation{%
  \institution{\ibm}
  \country{}
}

\author{Indranil Gupta}
\affiliation{%
  \institution{\uiuc}
  \country{}
}

\author{Saurabh Jha}
\authornote{Emails of authors: sc134@illinois.edu, frankeh@us.ibm.com, lshwart@us.ibm.com, chandras@us.ibm.com, indy@illinois.edu, saurabh.jha@ibm.com}
\affiliation{%
  \institution{\ibm}
  \country{}
}

\begin{abstract}
\noindent Research in compute resource management for cloud-native applications
is dominated by the problem of setting optimal {\it CPU limits} (\climit) -- a fundamental OS mechanism that strictly restricts a container's CPU usage to its specified \climit. Rightsizing and autoscaling works have innovated on allocation/scaling policies assuming the ubiquity and necessity of \climit. We question this. Practical experiences of cloud users indicate that \climit harms application performance and costs more than it helps. These observations are in contradiction to the conventional wisdom presented in both academic research and industry best practices. We argue that this indiscriminate adoption of \climits is driven by erroneous beliefs that \climits is essential for operational and safety purposes. We provide empirical evidence making a case for eschewing \climits completely from latency-sensitive applications. This prompts a fundamental rethinking of auto-scaling and billing paradigms and opens new research avenues. Finally, we highlight specific scenarios where \climits can be beneficial if used in a well-reasoned way (e.g. background jobs).

\end{abstract}

\setcopyright{none}
\settopmatter{printacmref=false}   
\maketitle


\vspace{-0.5em}
\section{Introduction} \label{sec:intro}

\begin{figure}[t]
\centering
    \includegraphics[width=0.7\linewidth]{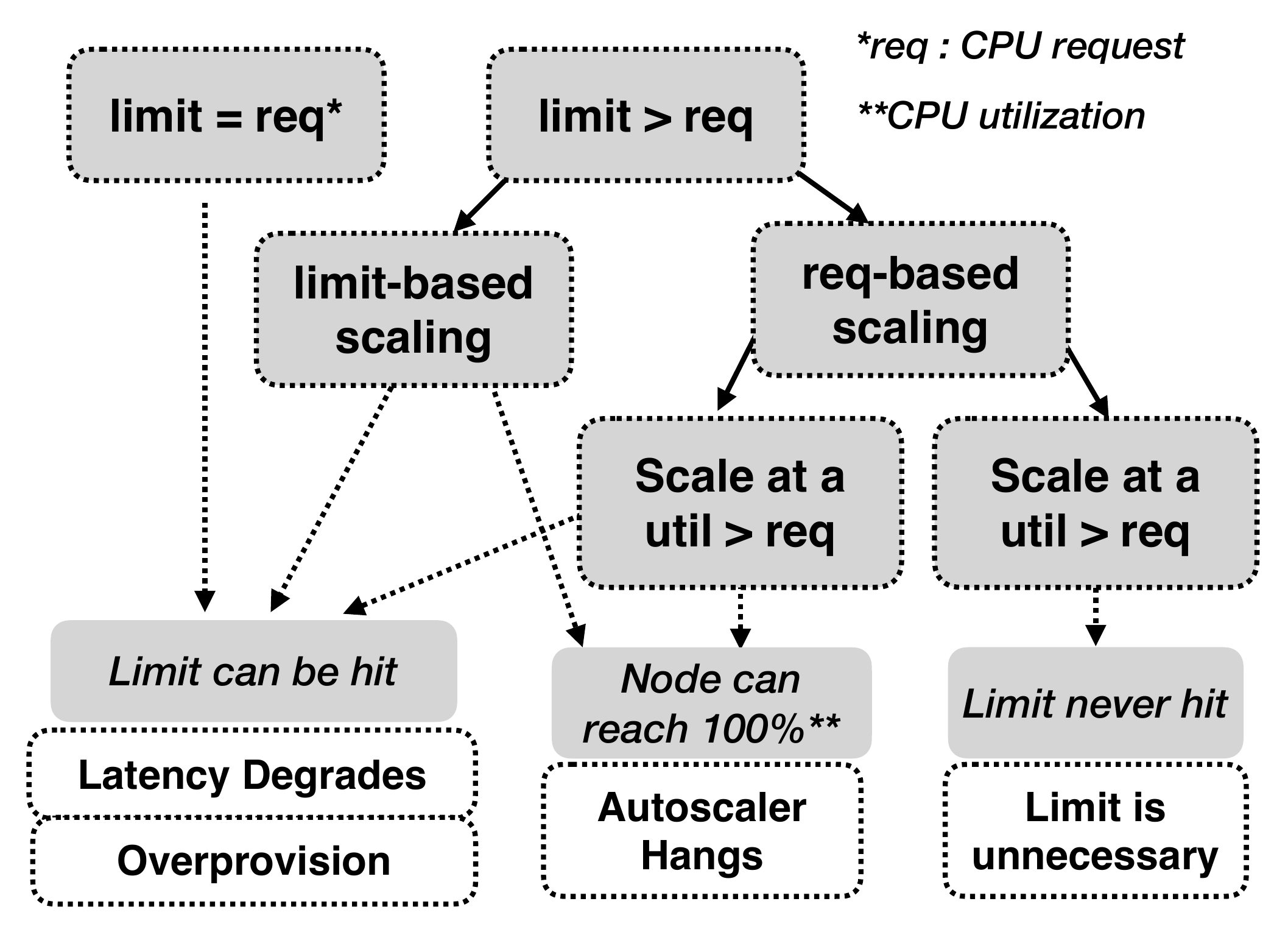}  
    \caption{\climits are either harmful or unnecessary: Summary of \S\ref{sec:motivation}.}
    \label{fig:motivation_summary}
\end{figure}

\begin{table*}[t]
  \centering
  \resizebox{0.95\linewidth}{!}{%
    \begin{tabular}{c|c|c|c|c|c|c|c}
      \toprule
      \rowcolor{white}
      {\it Properties} & FIRM~\cite{firm} & Cilantro~\cite{cilantro} & Autothrottle~\cite{wang2024autothrottle} & Ursa~\cite{ursa} & 
      SHOWAR~\cite{showar} & Erlang~\cite{sachidananda2024erlang}  & H(V)PA~\cite{hpa,vpa}\\
      \midrule
      {\it CPU alloc. mechanism} & \climits & \climits & \climits & \climits & \climits & VM*  & \crequest \\

      
      {\it Works without \climits?} & \xmark & \xmark & \xmark & \xmark & \xmark & \cmark & \cmark  \\

      \begin{tabular}{c} {\it What exactly fails} \\ {\it on disabling  \climits?} \textsuperscript{2}  \end{tabular}& \begin{tabular}{c} \S 3.4, Reward fn. $r_{t}$ \\ assumes $\frac{RU_i}{RLT_i} \leq 1$, else \\ ${RLT_i}=0$ trivially.  \end{tabular}
       & \begin{tabular}{c} \S 4.2, perf $p(a,l)$ is \\ f(alloc `$a$'), and usage $>$\\ alloc not considered. \end{tabular}
       & \begin{tabular}{c} \S 3.2.1 Uses throttle \\ count as the main \\ metric. NA w/o \climit. \end{tabular} &
       \begin{tabular}{c} \S MIP1 uses fixed  $R_i$   \\ for each LPR, obtained \\  by profiling w/ \climits.\\  \end{tabular}
       & 
       \begin{tabular}{c} \S 3.2 - 3.3: uses slack ($e$), \\ algo devolves to HPA\\ if slack is negative.  \end{tabular} &  NA
       & NA \\

      {\it \creq scaled with  \climit?}\textsuperscript{3} & \xmark \cite{firm_code_limit} & \xmark \cite{cilantro_code_limit} & \xmark \cite{autothrottle_code_limit} & NP**  & NP** & NA & NA \\

      {\it Container sizing technique} & Reinforcement Learning & Fixed & Contextual Bandit & Profiling \& MIP & Three-sigma rule & Fixed & Fixed(Manual) \\

      \bottomrule
    \end{tabular}%
    }
  \caption{Use of \climits in recent works on microservices CPU sizing/autoscaling. 
  *Erlang uses one node(VM) per pod replica. ~\textsuperscript{1}Horizontal scaling threshold manually set, not found in artifact. \textsuperscript{2}Relevant sections in that paper. \textsuperscript{3} Relevant code artifact cited. NP**: Code publicly not available. 
    }
  \label{table:related_work}
  \vspace{-0.5cm}
\end{table*}


The problem of finding optimal resource allocation for containerized microservices (\textit{rightsizing}) and automatically adjusting it with varying load (\textit{autoscaling}) has been a major focus of research and industry practitioners over the past decade. The objective of autoscaling research is to devise algorithms to meet the application Service Level Objectives (SLO) \cite{slo_definition} with minimal resource allocation ~\cite{cilantro, firm, wang2024autothrottle, sachidananda2024erlang, ursa, showar, pema}. These are built into orchestrators like Kubernetes (K8s)~\cite{kubernetes}, Borg~\cite{borg_wilkes}, Twine~\cite{tang2020twine}. They then handle the distribution of shared resources like CPU \& memory among the microservices. Management of CPU resource is of special interest since CPU usage closely correlates to request processing latency \& is often the primary resource bottleneck~\cite{ben_reids_cpu}.

We argue that these efforts are misguided, as the fundamental problem of CPU allocation for containers lies not in the policies ({\it how much} to allocate), but rather in the mechanism used today({\it how} it is allocated). {\it \cpulimits} (\clim in short) is a widely adopted mechanism, built on top of Linux's {\tt cpu.cfs\_quota\_us}~\cite{rightsizing_datadog, rightsizing_GKE, ursa, showar, cilantro}). Intuitively, \climit (typically specified in millicores) defines an upper bound on CPU usage beyond which the container is throttled. In the event of such throttling, application latency can dramatically degrade. Thus, "avoiding throttling" spawned academic autoscaling research focusing on algorithmic innovations to automatically adjust \climits (Tab. \ref{table:related_work}). Despite these efforts, the techniques, however, have seen low adoption in practice~\cite{datadog_report_1, datadog_report_2}. Practitioners started noticing that the impact of throttling on SLO and predictability outweighs the benefits of \climits, if any (\S\ref{sec:motivation}). So, we question, \textit{do we really need \climits?}


\climits have a history. Initially, on individual machines, users could control CPU allocation using \textit{shares} (\texttt{cpu.cfs\_shares}). The Linux CFS Scheduler (or recent EEVDF \cite{eevdf}) allocates CPU time to processes proportional to their specified share \cite{cpu_rehat, cgroups, ibm_cgroups}. \textit{Quota} (\texttt{cpu.cfs\_quota\_us}, aka bandwidth control) was introduced in Linux 3.2~\cite{quota_linux3.2} for two purposes:
\ding{182} To have a \textit{conceptually simpler} \& \textit{predictable} way of specifying CPU allocation, in contrast to \textit{share},  wherein CPU cycles a process gets is \textit{relative} to CPU use of all other co-located processes,  
and \ding{183} To support pay-per-use -- Operators needed a way to assign an explicit upper bound to CPU usage by tenants based on how much they paid. Subsequently, as containerization on cloud became popular, quotas \& shares got adopted for container sizing (as \clim \& \crequests in \ks terms).  

At first, needing \clim seems logical. When deploying applications in multi-tenant environments, we do not want containers to use excessive CPU, thereby depriving co-located containers of CPU resources, causing unpredictable latencies and potentially expensive cloud bills. Thus, having \clim was (and still is) considered a "best practice". For instance, in \ks, for critical containers to have the highest Quality of Service, 
they must have \clim (and equal to its \creq) \cite{pod_qos}, in spite of the risks of throttling.  

However, we raise two important considerations: First, use of \climits for \ding{182} is redundant today, because the orchestrators ensure the sum of \crequests of all containers placed on a node is less than the node's capacity~\cite{k8s_scheduler}. This makes \crequests an absolute minimum CPU guarantee(\S\ref{subsec:req_enough})  - we do not need \clim on top of \creq. Secondly, \climits is non-intuitive to set and has adverse impacts on \textit{all the three} key metrics -- application latency, reliability, and cost, offering no benefits (\S\ref{sec:motivation}, summarized in Fig.\ref{fig:motivation_summary}). So we posit that the community should not spend efforts on setting the right \climits, and provide directions on `what else if not \climit'. 


Our interviews with SREs (Site Reliability Engineers) revealed that the DevOps community is split too on the ``limits vs. no limits'' debate. Some insist CPU limits are essential to limit resource misuse, while others argue against it~\cite{stop_using_limits, use_limits, zalando}. In this paper, \begin{inparaenum}[(1)]
    \item we discuss both sides, and provide empirical evidence to eliminate \climits from latency sensitive application management.
    \item However, going \climits-free requires fundamental rethinking in existing autoscaling and billing paradigms. We elaborate research directions and demonstrate the potential by building a prototype autoscaler (\S\ref{sec:removing_limits}).
    \item Finally, judicious use of \climits in specific scenarios is warranted (e.g., background jobs). We highlight them while debunking several myths (\S\ref{sec:myths}) that promote the use of \climits.
\end{inparaenum}

\vspace{-0.5em}
\section{Background}
\label{sec:background}

Microservice applications typically are deployed as containers on a cluster of {\it nodes} (physical machines or VMs) running Linux. 
Container orchestration systems like \kubernetes (\ks)~\cite{kubernetes}, Borg~\cite{borg_wilkes}, Twine~\cite{tang2020twine}, etc. co-locate multiple containers on the same node, with Linux's Completely Fair Scheduler ({\it CFS}) on each node multiplexing CPU time among the containers based on their CPU allocations. 
We use the \ks term \textbf{`pod'} interchangeably with `container'.

\noindent{\bf CPU Shares and Quotas: } {\it CFS} gives 
processes a minimum CPU time proportional to their \textit{share} (\texttt{cpu.cfs\_shares}) and no more than their specified \textit{quota} (\texttt{cpu.cfs\_quota\_us})
~\cite{ibm_cgroups, google_limits, turner2010cpu}. A process's actual CPU time changes dynamically with more processes. 
In each scheduling period of 100ms, two processes $\mathcal{A, B}$ with 400 and 600 shares get a guaranteed 40ms ($\frac{400}{400+600} \times 100$ms) and 60ms of CPU time respectively. If $\mathcal{A}$ uses only 20ms of its allocated time, $\mathcal{B}$ is free to use the remaining 80ms. However, if $\mathcal{B}$ had a quota of 60ms,  it cannot exceed 60ms even if a residual 20ms is available. When $\mathcal{B}$ hits 60ms, it will be throttled until the next period.

\noindent{\bf \cpurequests and \textsc{limits}: }  For each container, the user can specify the CPU allocation using  \cpurequests and \cpulimits (we'll use \creq, \clim in short). Effectively, \textit{\textbf{\creq is the minimum guarantee and \clim is the maximum allowance on a container's CPU usage.}}  Millicores (m) units are used, where a millicore is 1ms worth of CPU time every second; so 1000m implies an allocation of one CPU core. Internally, \ks translates \creq and \clim to \texttt{cpu.cfs\_shares} and \texttt{cpu.cfs\_quota\_us} respectively. Importantly, \ks scheduler~\cite{kube-scheduler} ensures that the sum of \creq of all containers on a node is less than the node's net available CPU to prevent over-allocation and safeguard the \creq's guarantee. Container can use more than its \creq when free CPU cycles are available, but if its CPU utilization (\cpuutil) reaches \clim it will be throttled. Note \clim $\geq$ \creq and it is optional to set \clim.

\noindent{\bf Horizontal Pod Autoscaler (\hpa):} \hpa is \ks's native and most commonly used autoscaler~\cite{hpa}. To enable autoscaling, user sets a scaling threshold (\cputhresh). When the container's \cpuutil reaches a user-set scaling threshold (\cputhresh), \hpa adds another replica of the container with the same \creq. \cputhresh is usually specified as a percentage (e.g. 70\%) of the container's \creq.

\vspace{-0.5em}
\subsection{Prevalence of \cpulimits}

The key idea of using \clim to prevent containers from using more than the specified amount of resources is widespread. 

\noindent{\bf In Industry:} Examples of the use of \climits in industry:
\begin{enumerate}
    \item \ks assigns Quality of Service (QoS) classes to containers based on \climit. To have the highest QoS class, a container must have \climit set and must be equal to its \crequests \cite{k8s_qos}. 
    DevOps engineers thus add \climit on critical pods \cite{datadog_limit, qos_reddit, qos_redhat}.
    \item Multi-tenancy best practice considers  \climits (quota) as best practice \cite{multitenancy_limit1, multitenancy_limit2, multitenancy_limit3, multitenancy_limit4}.
    \item Popular software projects, in attempts to become `container friendly', use \climit to infer internal parameters like threadpool size, processor count etc \cite{rabbitmq_quota, erlang_quota, java_quota}.
    \item Managed services by cloud providers, like GCP's Autopilot \cite{rzadca2020autopilot, autopilot_reddit}, automatically apply \climits on customer pods. 
    \item Deployment templates in the widely used Helm Chart library Bitnami come with pre-set \climit values \cite{bitnami_limit}.

\end{enumerate}

\noindent{\bf In Research:} Significant research effort has gone into `finding optimal \climits to minimizes CPU allocations while meeting application SLO' ~\cite{firm, cilantro, showar, ursa, sachidananda2024erlang, rzadca2020autopilot}. 
\cs{Need more citations}
Table~\ref{table:related_work} lists the most recent autoscaling systems from top conferences. 
\cs{what other conf}
They rely on \climits to specify container CPU allocations. Tuning \climits is used as the primary mechanism of trading off CPU allocation for latency.

\vspace{-0.5em}
\section{\cpulimits Considered Harmful} 
\label{sec:motivation}

\begin{figure*}[t]
\centering
\hspace{-5mm}
\begin{subfigure}[b]{0.20\linewidth}
    \includegraphics[width=\textwidth] {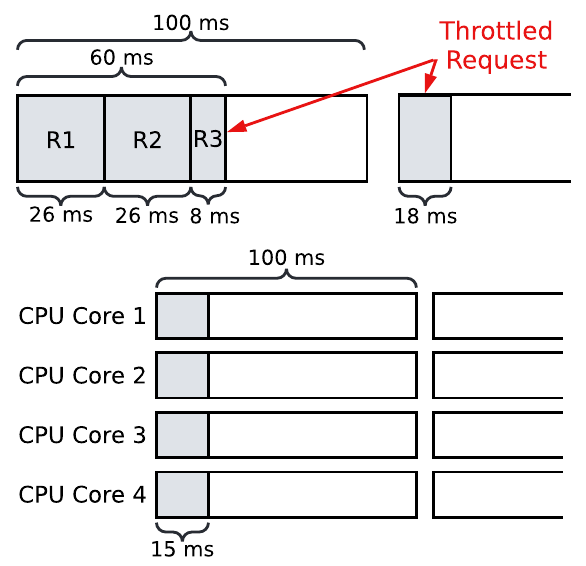}
    \vspace{-6mm}
    \caption{}
    \label{fig:throttling}
\end{subfigure}
\begin{subfigure}[b]{0.60\linewidth}
    \includegraphics[width=\linewidth]{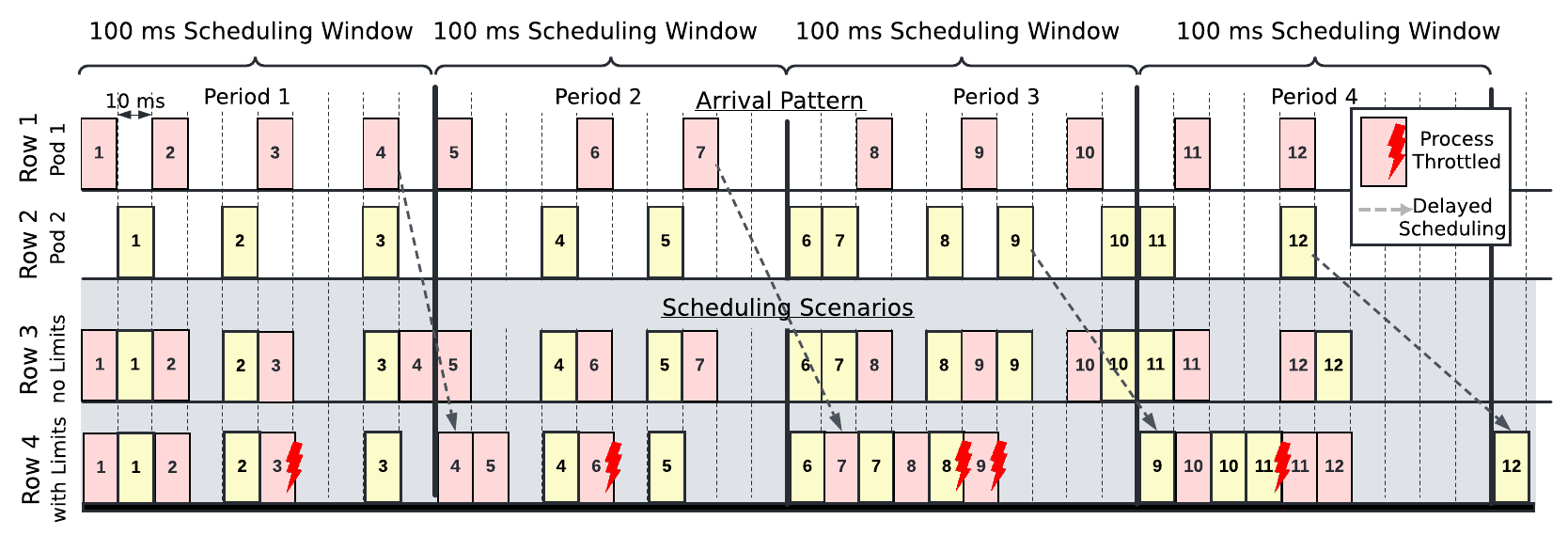}
    \vspace{-6mm}
    \caption{}
    \label{fig:motivation_queuing}
\end{subfigure}
\begin{subfigure}[b]{0.20\linewidth}
    \includegraphics[width=\linewidth]{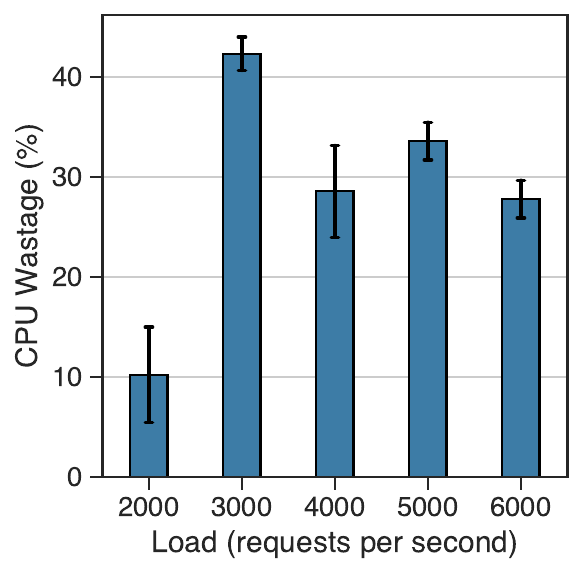}
    \vspace{-6mm}
    \caption{}
    \label{fig:hpa_with_without_limit}
\end{subfigure}
\caption{(a) Throttling of single-threaded and multi-threaded processes. (b) Formation of queues: Pods have \crequests = 300m millicore (m), i.e., 30ms per 100ms. In Row-4, \climit of 300 millicores is applied on both. We assume fair scheduling (with no preemption, for the sake of simplicity). (c)  Impact of \climits on cost: Percentage increase in CPU required to meet SLO with \climits specified using \climits vs just \crequest (SN app) }
\vspace{-5mm}
\label{fig:global2}
\end{figure*}

We now present evidence of the disconnect between industry standards, academic research, and practitioners' experiences with using \climits. We systematically argue for eliminating the use of \climits completely in latency-sensitive application deployments.

\vspace{-0.5em}
\subsection{The Ongoing Debate on Limits}
\label{subsec:limitsdebate}

This section is motivated by industry debates around the benefits of limits \cite{stop_using_limits, autopilot_reddit, cheuk_throttling_limits, tim_hockin_tweet, limits_hn_discussion, 73kpods, adobe_kubecon_talk}. 
Our study of online discussion forums reveals that cluster administrators are 
{\it already getting rid} of \climits~\cite{zalando, cheuk_throttling_limits}. 
We also interviewed several DevOps personnel at KubeCon (premier \ks conference) and inside a large cloud provider company. We found that admins would {\textit{largely prefer not having to set \climits}} due to frequent latency degradation from \climit's throttling.  
\hf{why, it has to come across that coming up with limits is hard and is often done very conservative leaving cycles idle, leading into next sentence with concrete evidence} 
%
%
Many use manual autoscaling, causing high management overheads and underutilization, e.g., {\it over 65\% of Kubernetes workloads use only 50\% requested CPU resources~\cite{datadog_report_2}.} 
We present select anonymized quotes from our interviews:

\squishlist
\item (Major Insurance Provider) ``\textit{A service with 16 CPUs didn’t utilize more than 10 CPUs due to throttling. We eliminated the issue \underline{by eliminating \climits altogether}.}'' 

\item (Major SaaS Enterprise) ``\textit{\underline{\climits is very non intuitive}, especially for multi-threaded examples. We are forced to use \climits...'' }
\item (Major Sports Company) ``\textit{(We) need a way for developers to not worry about resource requirements and CPU measure. We \underline{stopped using limits}; we mostly only use HPA.}'' 
\squishend


So, what is the {\it expected} benefit of using \climits?
We encountered three common reasons why DevOps personnel believe \climits are useful:  
\begin{inparaenum}[(1)]
\item for guaranteed CPU allocation \& predictable performance, 
    \item to provide safety \& cluster stability despite runaway containers that consume a lot of resources \cite{runaway_pod1}, especially in multi-tenant deployments \cite{multitenancy_limit1}, and
    \item to limit dollar costs for their application.
\end{inparaenum}
However, our measurements will show that \climits lead to unpredictable performance and cluster instability. Fig. \ref{fig:motivation_summary} is a decision tree that summarizes our arguments in the rest of this section. In all cases, the only way forward is to eliminate limits from autoscalers.

\noindent{\bf Setup: } We deploy HotelReservation and SocialNetwork applications from DeathStarBench~\cite{dtsb} on EC2 clusters. We optimize the CPU allocation vs latency over various \climits, \crequests, and scaling thresholds (\cputhresh) settings under varying loads. 

\vspace{-0.5em}
\subsection{\climits Exacerbate Latency \& Reliability} 
\label{subsec:limitsbadlatency}


\igc{You need a lead sentence about what this section is about.} \chirag{added}
First, we observe that using \climits to reduce CPU allocation by limiting the CPU utilization of pods can have adverse impacts.

\begin{figure*}[t]
\centering
\hspace{-5mm}
\begin{subfigure}[b]{0.33\linewidth}
    \includegraphics[width=\textwidth]{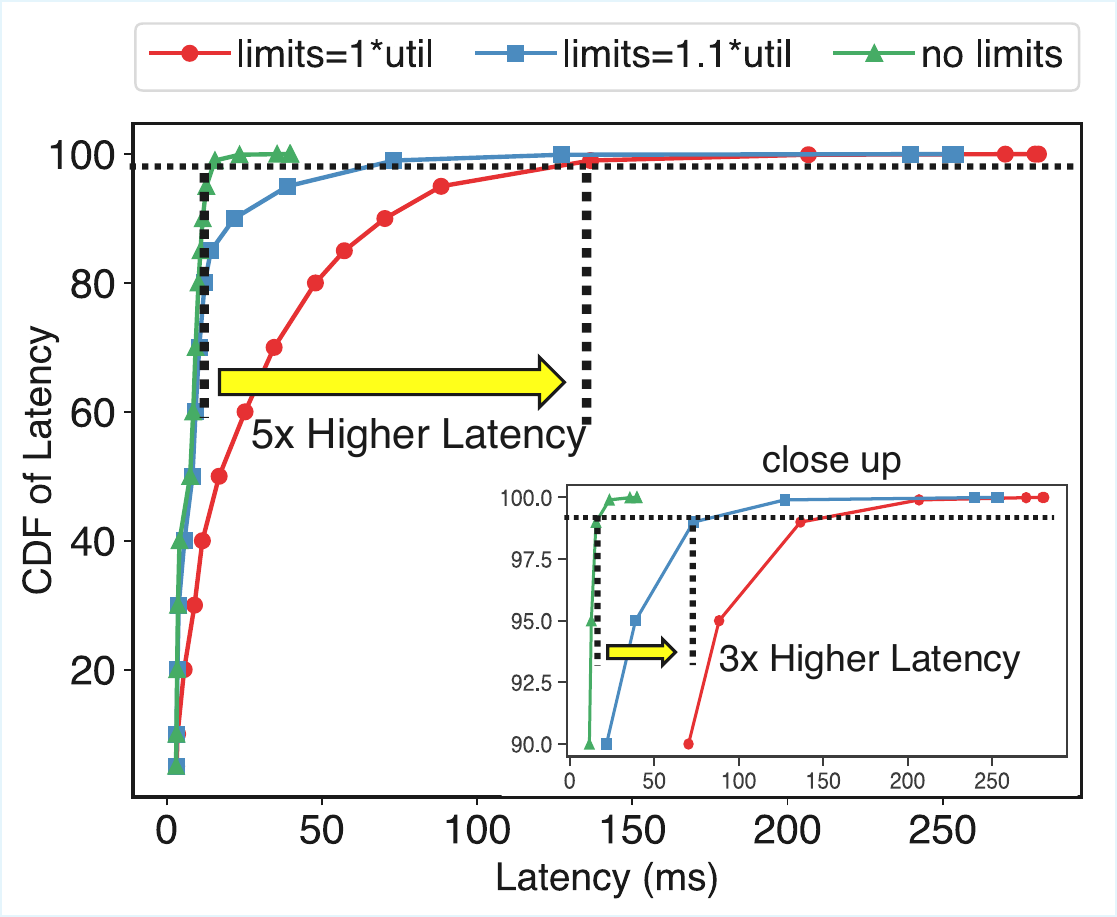}
    \vspace{-6mm}
    \caption{}
    \label{fig:cdf}
\end{subfigure}
\begin{subfigure}[b]{0.3\linewidth}
    \includegraphics[width=\linewidth] {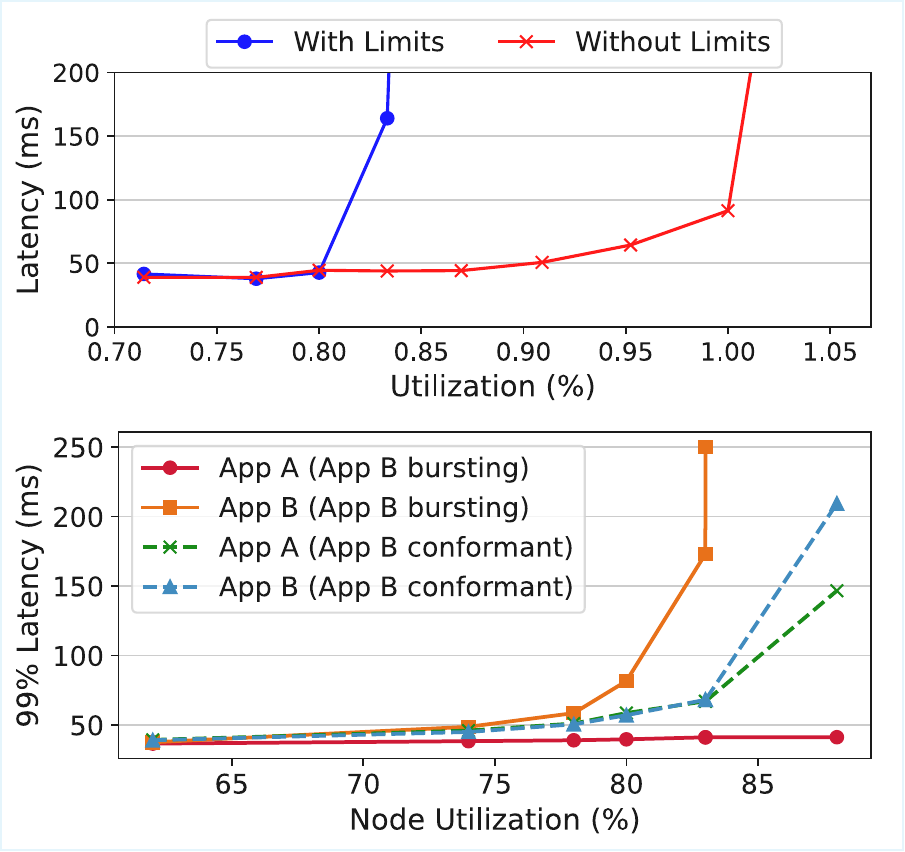}
    \vspace{-6mm}
    \caption{}
    \label{fig:request_is_enough}
\end{subfigure}
\begin{subfigure}[b]{0.33\linewidth}
    \includegraphics[width=\linewidth]{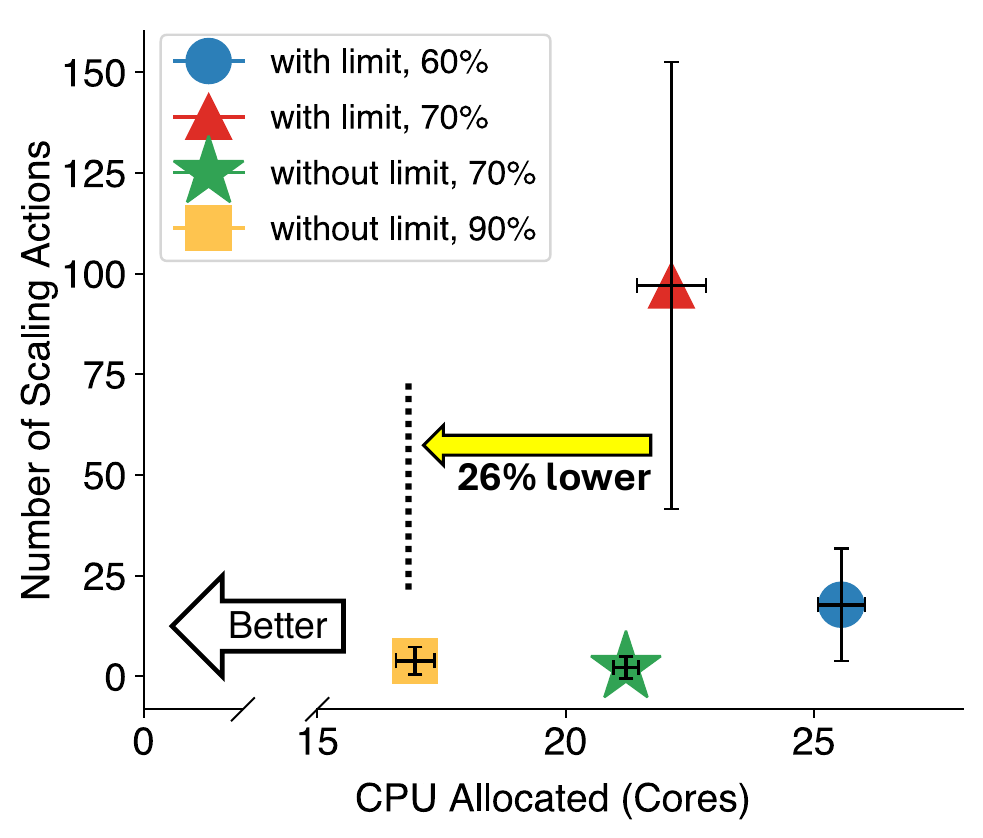}
    \vspace{-6mm}
    \caption{}
    \label{fig:timing_limits}
\end{subfigure}
\caption{(a) Impact on latency with \climits of 1\texttimes ~\& 1.1\texttimes ~the CPU util (HR). (b) [top] Queuing with and without \climits. X axis is \% utilization of pod's allocated \creq/\clim. [bottom] \creq protects app A against bursting app B  (c) No. of scaling actions \& CPU required to meet SLO on increasing load by 25\% different scaling thresholds (60\%, 70\%, 90\%) with \& without \climits (SN) 
(HR = HotelReservation, SN = SocialNetwork  \cite{dtsb}). \chirag{Change the para for unsafe hpa}  }
\vspace{-5mm}
\label{fig:global2}
\end{figure*}

\label{subsec:motivation_perf_impact} 
\noindent{\bf Bad for Latency: } 
For the same amount of CPU utilization of a pod, applying \climits on it results in worse latency characteristics.
\hf{rephrase, not clear}\cs{cleared after adding the into sentence?}
In Fig. \ref{fig:cdf}, we put \climit on only one (out of 19) pods of HotelReservation. Even \emph{without} attempting to reduce its actual CPU utilization, the end-to-end tail latency degraded by 5\texttimes.
\hf{Figure \ref{fig:cdf} suggests there is a Paretto curve and if one can show that the Paretto curve is equal no limit. You might need more than one app out DeathStar to avoid critics. Take 3 most popular one or all and make picture bigger. } \cs{I tried to get this but no time this time. Will add for full paper submission}

\climits harm application latency in two distinct ways:
\begin{inparaenum}[(1)]
    \item {\it throttling} of individual requests, and 
    \item formation of {\it long queues} across requests.  
\end{inparaenum}
Consider a pod with \climits of 600 millicores, i.e., upto 60ms of CPU time within every 100 ms scheduling period. Fig. \ref{fig:throttling} (top) shows a single-threaded container, where requests take $\sim$26 ms. Every one in three requests (e.g., R3) will be throttled~\cite{cpu_throttling} for 40ms, resulting in an execution time of 66ms. If the containers were multi-threaded, like Figure~\ref{fig:throttling} (bottom), the limit constrains the {\it cumulative} CPU time across all threads (e.g., avg. of 15ms CPU runtime per core with four threads). 

Further, throttling can form long queues as illustrated in Fig. \ref{fig:motivation_queuing}. Consider, two pods $P_{1}$ and $P_{2}$ co-located on a single-core node. Each pod gets a load with a Poisson arrival rate of 30 req/sec. Every request needs 10 ms of CPU time, translating to a net CPU utilization of 300 millicores (30ms per 100ms of scheduling period). We show two request execution schedules 
(1) when each pod is allocated 300 millicores \crequests (Row-3), and (2) when 300 millicores \climit is applied (Row-4). 
In both cases,  pod CPU utilization (30\%) and  net node utilization (60\%) are nearly the same, but average latency increases significantly with \climit. 
With \climits, typical microbursts of Poisson arrival form queues taking multiple scheduling periods to drain. E.g., in period 1, $P_{1}$ gets a microburst of 4 requests (instead of average of 3). Thus, request 4 gets delayed due to throttling, forming queues for requests 7 \& 10.

\textit{But isn't this regular queuing that is expected when job arrival rate matches the processing speed? } No. \climit-induced queuing is worse because (1) regular queuing happens only during high \textit{node} CPU utilization. \climit causes queueing at each \textit{pod}, and thus happens more often. (2) In regular queuing, a job in the queue waits for all the jobs in front to finish.  With \climit, in addition, 
the job waits for cumulative throttling times. Thus, queues form more rapidly and at lower loads in \climit-ed pods as shown in Fig. \ref{fig:request_is_enough} (top). 

\noindent{\bf Bad for Reliability: } In our experiments, tail latency degradations due to \climits occurred frequently and lasted multiple seconds--triggered even during routine cluster actions like scaling, updating deployments, or adding/removing nodes in the cluster.
Few cases resulted in cascading failures to other parts of the system. 
For instance, with long queue formation, the pod's memory consumption exceeded its memory limit, causing pod restarts.
This reduced the overall capacity and led to other pods hitting their limit or causing request timeouts.
Others have noted similar behavior \cite{limit_reliability_timeout, limit_reliability_memlimit}. 

\vspace{1mm}
\noindent{\emph{\textbf{Takeaway 1:} If a pod's CPU utilization hits \clim, latency drastically degrades \& can cause cascading failures. \clim creates a false allocation-vs-latency tradeoff, without actually reducing the \cpuutil.}}

\vspace{-0.5em}
\subsection{Why Not Set Higher \cpulimits?}
\label{subsec:motivation_deadlock}

A natural solution is to {\it always} ensure that the \climits is higher than the actual container utilization. However, we argue:

\vspace{1mm}
\noindent\textit{\textbf{ (1) If the \climit is ever hit, it causes severe latency degradation, and (2) If the \climit is never hit, then it is unnecessary.}}

\noindent{Two cases are possible depending on what we do with \crequests:}

\noindent{\bf Case A : \climits = \crequests  $\rightarrow$ Increased Cost:} 

To avoid throttling, we must maintain a {\it margin} between a pod's CPU utilization and \climit. Adding this margin to \crequest translates to larger pod footprints, which in turn require more nodes for placement since pods are packed according to their \crequest~\cite{kube-scheduler}. In Fig~\ref{fig:hpa_with_without_limit}, we manually optimized SocialNetwork to get the lowest \climit-based allocation while meeting an SLO. Further, to maintain this margin at all times, we configure \hpa with a scaling threshold(\cputhresh) of (100 minus margin). 

To maintain the tail latency, it took 25-45\% higher net \crequests than the overall observed CPU utilization. In other words, we had to increase the \climits by \textasciitilde 30\%, roughly split as a 10\% increase to handle Poisson microbursts, 10\% for fluctuations from cluster actions (eg: pod updates, scaling) and configurations (eg: child and parent colocated on a node ~\cite{lifting_fog_pod_colocation}), and 10\% for load variations (e.g. \cite{gmail_load_variation}). This also explains why the industry rule of thumb for margin is 20-30\%~\cite{hpa_threshold}. However, it varies by application, load levels etc--in Fig~\ref{fig:hpa_with_without_limit}, \texttt{nginx-thrift} pods used 30\%, while \texttt{user-timeline-service} needed 45 \%. 
Any attempts at reducing the margin (i.e, increasing scaling threshold) can lead to unpredictable, long convergence time to meet the SLO with varying load. In Fig. \ref{fig:timing_limits}, we increase the load on SocialNetwork by 25\% and measure the time taken to meet SLO. With a larger scaling threshold, CPU allocation reduces but leaves a smaller margin between the utilization and \clim (which causes throttling), trading it off with predictably meeting SLO (Fig~\ref{fig:timing_limits}). We observe a 4\texttimes increase in the time to meet SLO when the scaling threshold increases from 60\% to 70\%! This shows the sensitivity of autoscaling to the scaling threshold with \clim.

\noindent{\bf \climits are too fine-grained for Overall Dollar Costs: } Developers who want their application to stay within a dollar budget today need to grapple with {\it multiple} \climits of individual containers. This is cumbersome, error-prone, and 
{\it too fine-grained} for application-level goals. The need to maintain enough margins when applying \climits and the difficulties in doing so cost-effectively explain the complaints by system admins as in \S\ref{subsec:limitsdebate}. 
\cs{Unsurprisingly, this has been a major focus of autoscaling research.}

\vspace{1mm}
\noindent{\emph{\textbf{Takeaway 2:} Avoiding performance impact of \climits leads to over provisioning. Minimizing that cost is non-trivial}}.
\vspace{1mm}


\noindent{\bf Case B : \climits > \crequests $\rightarrow$ Autoscaler hangs}

Unlike Case A, if we increase only \clim without increasing \creq, we can prevent throttling while not increasing the cost. This is, in fact, a common practice~\cite{firm, wang2024autothrottle}. Depending on what we set \cputhresh to, either \clim will never be hit, rendering it unnecessary, or autoscaler may `hang,' leading to long periods of SLO violation. If \cputhresh is lower than \crequests--the autoscaler creates a replica before the container utilization reaches \crequests--the \clim will never be reached.
In this case, \climits serves no purpose. 

If \cputhresh is higher than \crequests, it scales when container utilization $>$ \crequests (e.g., all autoscalers in Tab. \ref{table:related_work}). But remember that a pod is only guaranteed its \crequest. Thus, if the pod is on a tightly packed node running at high utilization, the pod's utilization may never exceed its \crequest and thus not reach \cputhresh. Autoscaler will not add additional pod replicas, and hence, the SLO can remain violated indefinitely. \climits > \cputhresh > \crequests may thus lead autoscalers to get stuck.

\noindent{\emph{\textbf{Takeaway 3:} \climits higher than \crequests are unnecessary at best, and cause autoscaler to hang at worst. }}

\textbf{Conclusion:} No matter how \climits are applied on latency sensitive pods, it can only harm the performance, cost, and reliability. 

\climits is unnecessary and not the right tool for CPU resource management. Among the researchers who acknowledge this, a common refrain then is -- "\textit{But \climit is a `necessary evil' to deal with adverse \& failure scenarios like CPU hogging/bursting/runaway/buggy pods?}". Judicious use of \clim has valid use-cases, \& we detail it in \S\ref{sec:myths}. However, we argue that effective use of \crequests is adequate in most of these scenarios \& remains an unexplored direction. Use of \clim in fact should be an exception rather than the rule.

\cs{What about memory limits?}

\vspace{-0.5em}
\subsection{\crequests are sufficient}
\label{subsec:req_enough}


\crequest is sufficient to ensure a container gets its CPU allocation, irrespective of its co-located containers' CPU usage. It follows from the OS's fairness promise \& the orchestrator's gate-keeping :

\noindent{\bf CFS's Promise: } CFS \footnote{Or the recent EEVDF scheduler~\cite{eevdf}
} provides proportional fairness guarantee: If $n$ pods $\{P_{1}, P_{2}, ... P_{n}\}$ with \crequests of $\{r_{1}, r_{2}...r_{n}\}$ are running on a node with $C$ CPUs, CFS ensures that each $P_{i}$ gets a \textit{minimum} CPU time of $\frac{r_{i}}{\sum_{1}^{n} r_{j}} C$, no matter what the co-located pods are consuming~\cite{k8s_resource_management}. Internally, CFS uses \crequests to weigh the container's usage while deciding the scheduling priority. 

\noindent{\bf Orchestrator's Gate-keeping: } Placement algorithms in container orchestrators like \ks
ensure $\sum_{1}^{n} r_{j} \leq C$. \textit{Thus \crequests $r_{i}$ is the absolute minimum guarantee of CPU that $P_{i}$ will get.}

Consequently, CPU allocated using \creq is guaranteed even without using any \clim. \textbf{We call a pod \textit{conformant} if its \cpuutil $\leq$ \crequests, and as \textit{bursting} otherwise}. Pods are free to use CPU beyond their \creq allocation 
. But they can not `steal' CPU from conformant pods. A bursting pod can drive up the node utilization. This however, will only impact its latency. 


We demonstrate this in Figure \ref{fig:request_is_enough}(bottom). We co-locate two SocialNetwork apps $A$ \& $B$ on a 4 core node. App $A$ has a constant load and is conformant with \creq of 1100m. 
App $B$ has a growing load, but a fixed \creq of 700m. As the node CPU utilization increases due to bursting $B$, the latency of $B$ degrades. $A$ is not impacted. From $A$'s viewpoint, the node is only 45\% $\frac{(1100+700)}{4000}$ occupied at all times. Note however, that the node has spare \creq of 1200m ($= 4000-(1100+700)$). Thus, the orchestrator can pack more conformant pods on the node. Then $A$ may see increased latency with higher node utilization. Dotted lines shows the latency if $B$ were conformant as its load increased. Remember that this degradation would happen even if the conformant pods had \climits -- \climits would only degrade the latency further due to throttling.


\vspace{-0.5em}
\section{Beyond \climits: Problems \& Solution Sketch} 
\label{sec:removing_limits}

\begin{figure}[t]
\centering
\begin{minipage}[t]{0.40\textwidth}
\includegraphics[width=\linewidth]{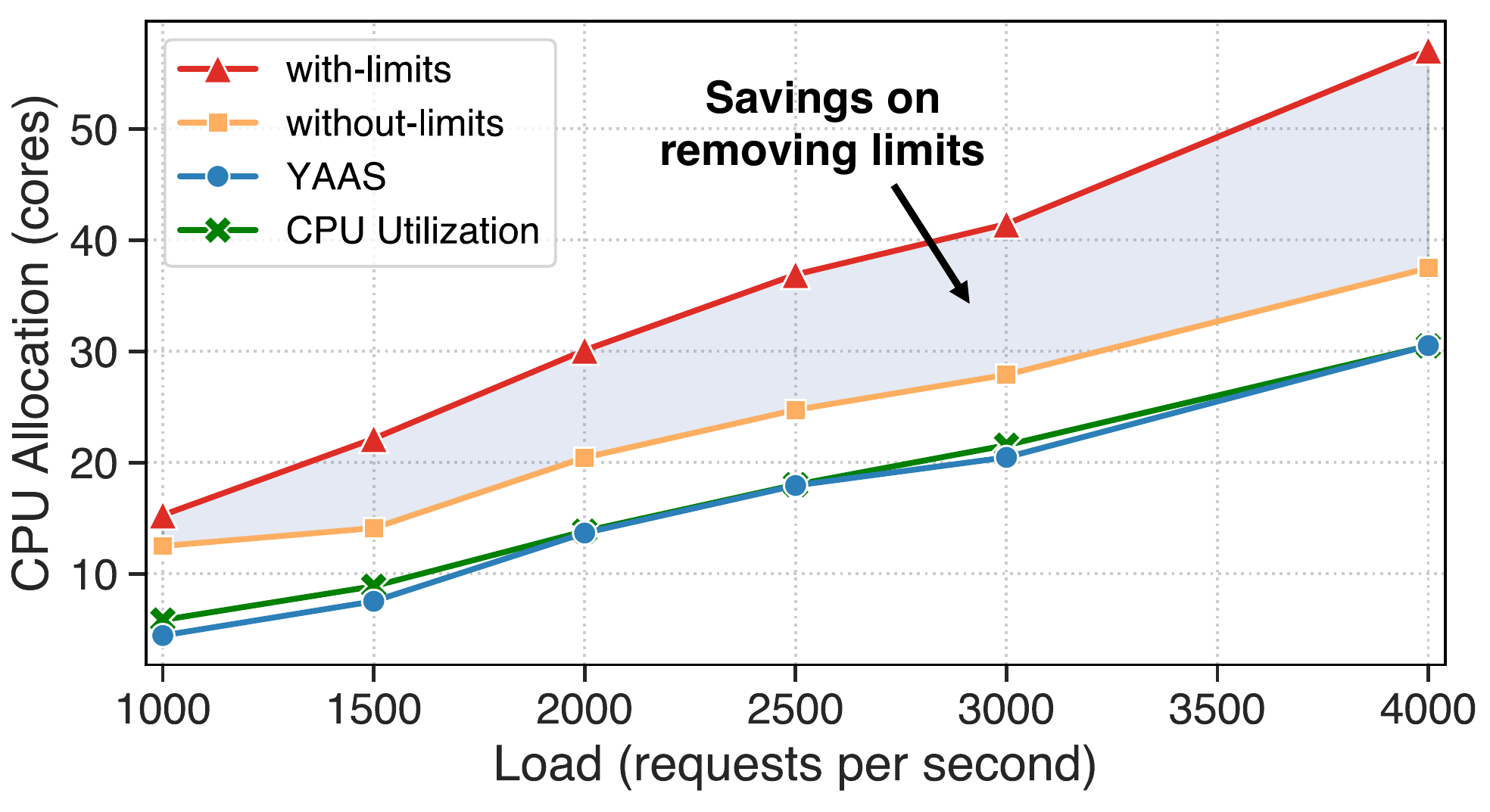}
     \caption{Savings on removing \clim \& with YAAS (HR app)}
    \label{fig:yaas_saving}
\end{minipage}

\label{fig:global1}
\end{figure}

Abandoning the use of \climits compels a course correction in CPU resource management practices. Here we call attention to two prominent areas that can benefit from fundamental rethinking: (1) Autoscaler designs, and (2) Billing paradigms. We build an illustrative prototype "Yet Another AutoScaler".

\vspace{-0.5em}
\subsection{Rethinking Autoscaler Design}
\label{subsec:rethink_autoscaler}
\cs{Idea:--- Trading CPU requests among pods -- \crequest relates to both latency and resource allocation}

Today, the concept of \climits is inherent in the modeling of autoscaling as an optimization problem. Removing limits renders many past works inapplicable for \climit-less design (Tab. \ref{table:related_work}). On the other hand, new opportunities arise once we shift focus away from  "avoiding throttling". 
In this section, we answer three questions: \textit{Why not simply replace \climits with \crequest? Without \climit to control CPU usage of containers, what does the latency-resource tradeoff look like? 
Finally, is all this effort in redesigning worth it?}

\noindent{\bf Current autoscaling models break with \crequest: } 
Autoscaling systems heavily used \clim because \clim provides an useful invariant : \textit{for any container $C$, the actual CPU utilization ($U$) is less than the \climit ($L$) at all times}. With \clim, as $U$ approaches $L$, the latency degrades due to throttling. Thus autoscaling can be conveniently formulated as "minimizing the slack ($L - U$) while keeping the latency under the SLO". $L$ was the knob used to tradeoff CPU allocation and application latency. Allocation minimization algorithms were designed, assuming the presence of \climits.

On removing \clim however, this model breaks: $U$ is no longer bound by the \crequest-based allocation $R$. For instance, in FIRM, the reward function used by Reinforcement Learning (RL) agent is $\alpha(\frac{SLO}{curr\_latency}) + (1-\alpha)(\frac{U}{L})$. With \clim, $\frac{U}{L}<1$ always. On replacing \clim($L$) with \creq ($R$), pod can burst and  $\frac{U}{R} > 1$ is possible. Thus the RL will converge to degenerate solution of setting $R = 0$ to maximize the reward 
(more failure examples in Tab.~\ref{table:related_work}).
With \clim, throttling dominated the latency vs CPU allocation tradeoff. Impact of node utilization 
\& \creq values were secondary (e.g. \cite{firm, wang2024autothrottle} in Tab. \ref{table:related_work} solely optimized \clim without modifying \creq), but will now come forth and must be actively managed.

\noindent{\bf The \climits-less Design Space: } Firstly, we need a way to trade off CPU allocation for latency of a pod. There are two such \textit{knobs}.
\begin{inparaenum}[(1)]
    \item \textbf{Overage $(U-R)$} of a pod's utilization compared to its allocated \creq, 
    \item \textbf{Net node CPU utilization ($N$)} of the node on which the pod is placed .
\end{inparaenum}
As shown in Fig.~\ref{fig:request_is_enough} (and \S\ref{subsec:req_enough}), having a smaller \creq on a pod (i.e. a higher overage) will sacrifice latency for resource and vice versa. Similarly, placing the pod on nodes with higher $N$ (i.e. tighter packing) degrades its latency. Fortunately, the latency varies smoothly on tuning these knobs unlike with \clim (ref. Fig.~\ref{fig:request_is_enough}(top)). This opens up new design opportunities for adaptive algorithms. 

\begin{figure}
    \centering
    \includegraphics[width=0.95\linewidth]{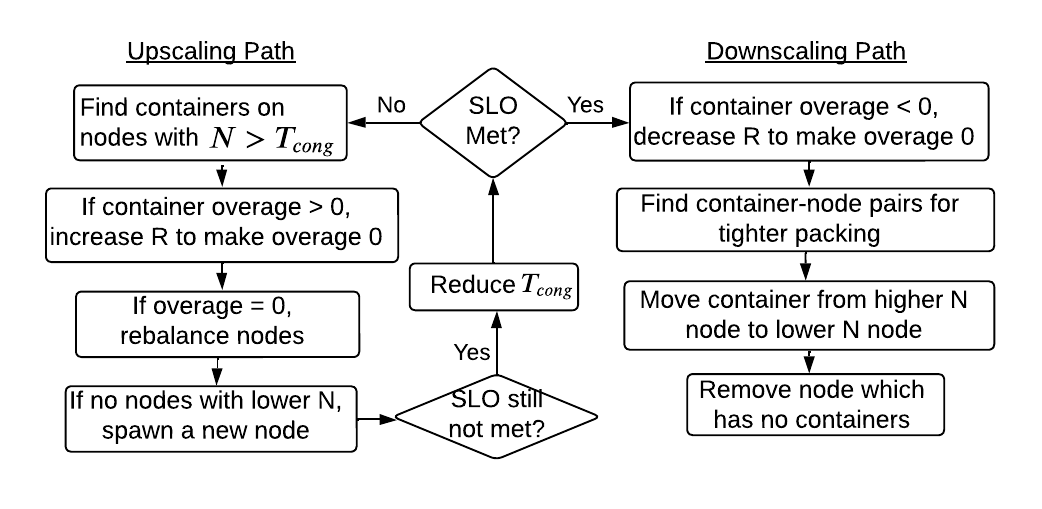}
    \caption{YAAS's scaling policies.}
    \label{fig:yaas_flowchart}
\end{figure}

To grasp the rich optimization space, consider the basic problem of finding smallest node that fits two pods C1 and C2. Suppose C1 needs an SLO of 100ms and has an average CPU utilization of 2000m. C2 has a 25ms SLO with 1000m utilization. One simple solution is to fix the overage at 0 i.e. set \creq of each pod equal to their utilization. Then we can search what node utilization ($N$) can meet the SLO of C2, the stricter of the two (stricter the SLO, lower the node util. has to be). Let's say a 4 core node (4000m) at 75\% ($\frac{2000m+1000m}{4000m}$) node utilization meets C2's SLO. However, with that large 25\% margin C1's SLO may be met too well ($<<$100ms). We can optimize further by tuning overages. Redistributing \creq to be say 1500m for both C1 , C2 i.e ($-500m$ \& $500m$ overage respectively), we can possibly place them on a smaller 3 core node at 100\% node utilization. We built YAAS to
show the potential of \clim-less designs.

\noindent{\bf A simple `Yet Another Autoscaler' (YAAS): } YAAS optimizes allocation by keeping overage $\geq 0$ for all containers at all times. We built YAAS atop \ks's \hpa and made simple optimizations. We let the overage grow until SLO is slightly violated, and then set the overage to 0 by 
allocating more resources (either by increasing \creq or by spawning new replicas).
Furthermore, when a new replica is added, the \creq of each replica is reduced proportionally and split equally across all active replicas to maintain overage $\geq 0$ and save resources compared to \hpa. This is because the utilization $U$ is split between the replicas. 
To reduce the node utilization $N$, YAAS moves containers across nodes if a container is scheduled on a {\it congested node}--whose utilization crosses a threshold ($T_{cong}$). Fig.\ref{fig:yaas_flowchart} shows the full operation. Fig.\ref{fig:yaas_saving} shows the CPU resources required to meet SLO (20ms) of the HotelReservation benchmark. \textit{Simply removing \climits achieves an average 38\% savings, and applying the above optimizations with YAAS further increases savings to 51\%.}

Several improvements are possible. In doing so, the designers must consider the following: Firstly, the two knobs are not independent. The same overage will result in different latencies on nodes with different $N$. On a node with a higher $N$, latency is more sensitive to overage. Secondly, vertical scaling is easier in the \climit-free domain. Pods are free to use spare CPU cycles on the node as the load fluctuates. With no threat of throttling, vertical scaling no longer requires the autoscaler to micro-manage the CPU allocations during every small load change. Thirdly, combining pod scaling with node scaling becomes necessary when using $N$ as one of the knobs. This differs from the status quo where pod scalers like \ks HPA or those in Tab.~\ref{table:related_work} work independently of node scalers like ClusterAutoscaler~\cite{cluster_autoscaling}, Karpenter~\cite{karpenter}, etc. Finally, safeguards such as hysteresis are needed to ensure the stability of multiple interacting control loops \& upscaling-downscaling logic.

\vspace{-0.5em}
\subsection{\clim-free Cloud Billing }

\climit is an essential mechanism for usage-based billing in multi-tenant cloud services~\cite{oracle_multitenancy, k8s_multitenancy}. \clim make billing easy for operators by offloading the responsibility of rightsizing to the users. Users are billed for the resource they ask for. We call this \textit{`resource-based billing'}. Fair \clim-based billing is an active area of research~\cite{ppu_nsdi}. Here we specifically ask how to support billing without using \clim?

\noindent{\bf Resource-based billing does not work \climits-free:} Consider an obvious approach to bill based on the \creq users specify. This has two issues: (1) the operator can not police the users' pods without \clim. Automatically adding \clim (=\creq) can escalate latency/cost for the users (e.g. \cite{autopilot_reddit}), (2) It is easy to exploit -- user can launch many pods with very small \creq, landing a few in emptier nodes. Once there, the pods burst out using more than they are billed for. Borg reported having this issue which they fixed with placement heuristics \cite{wilkes_goto_talk}.
Alternatively, we could bill by the actual \cpuutil of the pods. But this can lead to large increase in user bills. Not all \cpuutil contributes to useful performance. For instance, containers that rely on spinning (e.g., based on Erlang~\cite{erlang_cpu_hog, erlang_limit_issue} or Go~\cite{elixir_go}) can exhibit high \cpuutil without proportional performance gains. Further, contentions such as the cache can increases pod \cpuutil (processes spend longer on the CPU due to cache misses). Users will end up paying more for worse performance. 

Hence, we propose \textbf{Performance Based Billing}. Users only specify what performance they desire \& a maximum cost limit - e.g: maintain SLO $\leq$ 100ms with net cost under \$XXX. Operators are free to manage resources as needed. With controls over node utilizations, pod placements, observability etc, operators have significantly more knobs to control the latency of applications. We show how such a scheme can be built on top of a \clim-free autoscaler using YAAS.

\noindent{\bf A sketch of \clim-free Billing: }
Recall that YAAS (\S\ref{subsec:rethink_autoscaler}) increases \creq conservatively (overage $\geq$ 0) -- only when the SLO is about to be breached -- and decreases it eagerly (Fig.~\ref{fig:yaas_flowchart}). Thus \creq set by YAAS is indicative of the minimum \creq needed to meet the SLO. We can use sum of \creq set by YAAS to bill the user. This however is susceptible to an adversarial attack: the user can set a very strict SLO. In an attempt to meet the SLO, YAAS will move the pod to lightly-loaded nodes (low $T_{cong}$) thus enjoying superior latency without paying for it.
To plug this gap, pricing should also reflect the node utilization $N$ the pod needs to meet SLO. The lower the node utilization that is needed, the higher the price per \creq.

\vspace{-0.5em}
\section{Debunking Myths \& Valid \clim Usecases}
\label{sec:myths}

Despite the evidences \& opportunities presented thus far, researchers and SREs alike expressed skepticism about abandoning \clim. We address them here.
Use of \clim is driven by erroneous beliefs that \clim is essential for operational and safety purposes. In reality, \crequest is largely sufficient. \clim must be used judiciously only in specific scenarios. We elaborate.

\noindent{\underline{\bf Myth 1: Multi-tenancy:} \textit{ \climits is needed in multi-tenant clusters to ensure applications get their assigned allocation}}

\noindent{\bf Reality:} False. \crequest is sufficient (\S\ref{subsec:req_enough}) 

\noindent{\bf Valid Usecase: Benchmarking}: Latency with \clim is a worst-case estimate of application's performance. Such benchmarking can help capacity planning. However,\climits should be limited to offline profiling and not be used in production.

\noindent{\underline{\bf Myth 2: Performance Isolation:} \textit{\climits provides performance isolation. It is a protection against noisy neighbors.}}

\noindent{\bf Reality:} False, with Caveats.
Co-located containers can affect each other through CPU resources in two ways: 1) High node utilization: \creq already provides a level of performance isolation by preventing bursting containers from affecting neighbors. \clim has no additional benefit.
2) Contention on micro-architectural resources, primarily cache \& memory bandwidth (MB). Effectively managing these interferences requires hardware support like Intel's CAT, MBA~\cite{cat, mba}. \clim does not provide these.

\noindent{\textbf{Valid Usecase: Background (BG) Jobs: }}
In the absence of CAT/MBA, \clim is often useful to control non-latency sensitive, throughput-oriented jobs like ML~\cite{limits_ml_job}, or momentarily bursty events like GC processes which exhaust the MB/cache~\cite{fried2020caladan}. First step is to set the lowest \creq (1m) for the BG jobs or use lower scheduling priority settings like SCHED\_IDLE~\cite{sched_idle}. But it may not be enough \cite{heracles}, and we may need to explicitly limit the BG jobs using \clim.

\noindent{\textbf{Valid Usecase: CPU Pinning: }} Some workloads benefit from being pinned to CPUs~\cite{mxfaas, heracles}. \clim has recently been modified to allow pinning if \clim is an integer~\cite{node_cpu_management}. 
But pinning has a negative impact on a highly threaded application \cite{chen2017workload}. Also, rounding to an integer can accumulate into significant wastage (e.g: as noted in Borg~\cite{wilkes_goto_talk})

\noindent{\underline{\bf Myth 3: Safety:} \textit{Runaway/buggy pods may consume excessive CPU. \climits is a must to contain such behavior \& control cloud bill}.}

\noindent{\bf Reality:} False. \clim can not be configured to prevent runaway pods. \clim has no way to distinguish `useful' CPU consumption (e.g., due to increased workload) from a bug-induced one.
Two scenarios: if autoscaling is enabled on that pod, then as \cpuutil approaches \clim, the autoscaler will scale up the pod, irrespective of whether the container is buggy or not. If autoscaler is not enabled, then the impact is limited to one node, and \creq protects co-located containers. Further, to limit the cloud bill, a \textit{full-application level limit} suffices~\cite{namespace_limit}. Applying \clim on every pod is overkill.

\noindent{\underline{\bf Myth 4: Cluster Stability:} \textit{Critical system components will become unresponsive at very high node utilization without \climits}.}

\noindent{\bf Reality:} False. Small amount of \creq must be reserved on each node for system components~\cite{reserve_cpu_system_daemons} 
\creq will guarantee the system pods gets CPU time no matter what the node utilization is.

\noindent{\bf Other Valid Use Cases:} Power capping~\cite{power_cap_osdi}.
Sometimes, throttling the containers is intentional in using \climits. E.g., we spoke to a company that hosts Minecraft servers, and they want their free-tier users to see their CPU utilization and throttle when they hit the quota (thus prompting them to upgrade).

\noindent{\bf Concluding remarks:}

In this paper, we showed that the use of \clim can be counterproductive and we recommend removing \clim from CPU resource management for latency-sensitive applications. We can and should base new autoscaling and billing systems on \crequest only. The design space here is rich, and even simple policies have potential for large resource savings, while being more reliable and predictable. \clim must only be used in specific, unavoidable scenarios.



\newpage
\bibliographystyle{plain}
\bibliography{main}



\end{document}